\documentclass[11pt]{article}
\usepackage{amsmath, amssymb}

\newcommand{\vct}[1]{\boldsymbol{#1}}
\newcommand{\ket}[1]{|#1\rangle}
\newcommand{\bra}[1]{\langle #1|}

\newcommand{\mtrix}[3]{\langle \,#1\,|\,#2\,|\,#3\,\rangle}
\newcommand{\kf}{k_{\textrm{F}}}

\newcommand{\snbar}{S_{\bar{\textrm{N}}}} 
\newcommand{\spnbar}{S_{\textrm{p}\bar{\textrm{N}}}} 
\newcommand{\snbarh}{S_{\bar{\textrm{N}}\textrm{h}}} 
\newcommand{\snn}{S_{\textrm{N}}}
\newcommand{\snosea}{S_{\textrm{NoSA}}}
\newcommand{\sph}{S_{\textrm{ph}}}
\newcommand{\svac}{S_{\textrm{vac}}}

\begin{document}

\baselineskip=20pt

\noindent
{\LARGE\textbf{%
Commutators of the four-current and sum rules in relativistic nuclear models
}}

\baselineskip=15pt

\vspace{\baselineskip}

\noindent
Haruki Kurasawa$^{1,2}$ and Toshio Suzuki$^3$

\vspace{0.5\baselineskip}

\noindent
{\small% 
$^1$Department of Physics, Graduate School of Science,
Chiba University, Chiba 263-8522, Japan\\
$^2$Center for General Education, Chiba University, Chiba 263-8522, Japan\\
$^3$Fukui Study Center, The Open University of Japan, AOSSA,
Teyose 1-4-1, Fukui 910-0858, Japan 
}

\vspace{\baselineskip}

\noindent
{\small
%\begin{abstract}%
There is a long-standing problem on the linearly energy-weighted
sum of the excitation strengths in the relativistic field theory
and nuclear models: The sum value should be positively definite,
while its naive calculation using the current commutator or the double
commutator of the excitation operator with Dirac Hamiltonian yields
the value to vanish. This paradoxical contradiction is solved
in an analytic way.
}

\section{Introduction}\label{intro}

The linearly energy-weighted sum $S$ of the excitation strengths is expressed
with use of the double commutator of the excitation operator $F$ with
the Hamiltonian $H$,
\begin{equation}
S = \sum_{n}(E_n-E_0)|\mtrix{n}{F}{0}|^2
= \frac{1}{2}\mtrix{0}{[F^\dagger,\ [H,F\ ]\ ]}{0},\label{s}
\end{equation}
where the closure property, $\sum_n\ket{n}\bra{n}=1$, is employed,
$\ket{n}$ denoting the eigen-state of $H$, $H\ket{n} = E_n\ket{n}$.
If there exist excited states with $(E_n - E_0) > 0$ and
$\mtrix{n}{F}{0}\neq 0$, the value of the sum should be obviously positive.

In non-relativistic models, for example, the well known f-sum rule value
$S_{\textrm{f}}$
is obtained for a $A$ particle system as\cite{suzu},  
\begin{equation}
S_{\textrm{f}}= \frac{A}{2m}\vct{q}^2,\label{fsum}
\end{equation}
for
\begin{equation}
F=\sum_{i=1}^Af(\vct{x}_i), \qquad
f(\vct{x}_i)=\exp(i\vct{q}\cdot\vct{x}_i), \label{fx}
\end{equation}
since the double commutator becomes to be a constant,
\begin{equation}
[F^\dagger,\  [\ H,F\ ]\ ]
=
\sum_{i=1}^A
\ [\ f^\ast(\vct{x}_i), \ [\ \frac{\vct{p}^2_i}{2m},\  f(\vct{x}_i)\ ]\ ]
=
\frac{A}{m}\vct{q}^2.
\end{equation}
Here, the Hamiltonian is assumed to be
\begin{equation}
H = \sum_{i=1}^Ah_i,\quad h_i=\frac{\vct{p}^2_i}{2m}+V(\vct{x}_i),
\end{equation}
with the potential $V(\vct{x}_i)$ which commutes with $F$.

In relativistic models, however, Dirac Hamiltonian
 contains the first derivative only,
\begin{equation}
 h= \vct{\alpha}\cdot \vct{p} + \beta m + V(\vct{x}) ,
\end{equation}
so that the the double commutator vanishes,
\begin{equation}
[ \  f^\ast(\vct{x}), \  [\  h, \  f(\vct{x}) ]\  ] = 0 , \label{dirac}
\end{equation}
in contradiction with $S>0$.

Let us briefly review more generally the above result of the relativistic
case according to the field theory.
The nuclear four-current is given in terms of the nucleon
field $\psi(\vct{x})$ by,
\begin{equation}
J^\mu(\vct{x})=\overline{\psi}(\vct{x})\gamma^\mu\psi(\vct{x}).
\label{fourcurrent}
\end{equation}
Since the excitation operator $F$ with a function $f(\vct{x})$
is defined as  
\begin{equation}
F = \int d^3x\,f(\vct{x})J^0(\vct{x}),
\end{equation}
the current-conservation
\begin{equation}
[\ H\ ,\ J^0(\vct{x})\ ] = i\vct{\nabla}\cdot\vct{J}(\vct{x})
\label{currentcons}
\end{equation}
provides us with
\begin{equation}
[\ F^\dagger,[\ H, \ F\ ]\ ] 
= \int d^3xd^3y\, f^\ast(\vct{x})f(\vct{y})
\,i\vct{\nabla}_{\vct{y}}\cdot[\ J^0(\vct{x}),\ \vct{J}(\vct{y})\ ].
\label{com} 
\end{equation}
In using the the anti-commutation relation as usual,
\begin{equation}
 \{\ \psi_m(\vct{x}),\ \psi^\dagger_n(\vct{y})\ \}
  =\delta_{mn}\delta(\vct{x}-\vct{y}),\label{field}
\end{equation}
$m$ and $n$ being the Dirac matrix indices, the nuclear four-current satisfies
\begin{equation}
[J^\mu(\vct{x}), \ J^\nu(\vct{y})\ ]
 = \psi^\dagger(\vct{x})
[\ \gamma^0\gamma^\mu,\ \gamma^0\gamma^\nu\ ]\psi(\vct{x})
 \delta(\vct{x} -\vct{y}).
\end{equation}
Thus, the time-component $J^0(\vct{x})$ and the space-component
$\vct{J}(\vct{y})$ of the current commute with each other,
\begin{equation}
[\ J^0(\vct{x}),\ \vct{J}(\vct{y})\ ] = 0.\label{current}
 \end{equation}
This fact makes Eq.(\ref{com}) vanish, in contradiction with $S>0$.

In the nonrelativistic framework, the commutation relation corresponding
to Eq.(\ref{current}) is written as\cite{suzu},
\begin{equation}
[J^0(\vct{x}),\ \vct{J}(\vct{y})]
=
-\frac{i}{m}\sum_{k=1}^A\delta(\vct{y}-\vct{x}_k)\vct{\nabla}_{\vct{x}}
\delta(\vct{x}-\vct{y}),\label{clcurrent}
\end{equation}
with the nonrelativistic four-current:
\begin{equation}
J^0(\vct{x}) = \sum_{k=1}^A\delta(\vct{x}-\vct{x}_k),\qquad 
\vct{J}(\vct{x})
=\frac{1}{2m}\sum_{k=1}^A\{\ \vct{p}_k,\ \delta(\vct{x}-\vct{x}_k)\ \}.
\end{equation}
By inserting Eq.(\ref{clcurrent}) into Eq.(\ref{com}) and using
\begin{equation}
\mtrix{0}{i[J^0(\vct{x}),\ \vct{J}(\vct{y})]}{0}= \frac{1}{m}\rho(\vct{y})
\vct{\nabla}_{\vct{x}} \delta(\vct{x}-\vct{y}),\quad
\rho(\vct{x})=\mtrix{0}{\sum_{k=1}^A\delta(\vct{x}-\vct{x}_k)}{0},
\label{eclcurrent}
\end{equation}
we obtain from Eq.(\ref{s}) \cite{suzu} 
\begin{equation}
S=\frac{1}{2m}\int d^3x\,\rho(\vct{x})|\vct{\nabla}f(\vct{x})|^2,
\label{sff}
\end{equation}
where $\rho(\vct{x})$ stands for the ground state density of
the many-body system. If we set  $f(\vct{x})=\exp(i\vct{q}\cdot\vct{x})$
in Eq.(\ref{sff}), we have the f-sum rule Eq.(\ref{fsum}).

For the last more than 50 years, much has been written, from
different points of view, on the above problem in relativistic
field theory\cite{goto,sch,itz,jac,g,w,na} and nuclear models\cite{mc}.

 In the relativistic field theory, Schwinger\cite{sch} pointed out
that Eq.(\ref{current}) should have a gradient of a $\delta$-function
on the right hand side from Lorentz covariance considerations\cite{jac}.
That additional term called Schwinger term plays an important role,
especially in current algebra, and is widely
explored\cite{itz,jac,g,w,na}, but its form is not well defined yet.
For example, on the one hand, Schwinger reproduced the term by introducing
the point-split current\cite{sch}.
On the other hand, Gasiorowicz and Geffen\cite{g} derived it by using 
the vector-meson dominance model,
while Weinberg and Gross et al.\cite{w} discussed it with use of
SU(3)$\times$SU(3) algebra.

In the nuclear study, Price et al. \cite{mc} tried to
interpret Eq.(\ref{dirac}) by invoking, in addition to usual
particle-hole excitations, transitions of particles in Fermi sea
to negative energy states in Walecka-Serot model\cite{sw,ring}.
The reason of the contradiction, however, is not made clear,
and a role of Schwinger term in this nuclear model has not been
investigated so far.
 
The purpose of the present paper is to show that within a framework
of the local field theory, the correct calculation of the right hand side
of Eq.(\ref{s}) yields  Schwinger term which is responsible for
a positive value of $S$.

In the following section, we will define the relativistic four-current
in the finite momentum space, where the time- and space-components
do not commute with each other. If we make the space infinite,
the commutator will disappear as in Eq.(\ref{current}),
yielding the contradiction.
In \S\ref{exp}, however, it will be shown that the expectation value
of the commutator should be calculated first, keeping the momentum space
to be finite. That expectation value does not vanish, even in letting
the momentum space be infinite later. The relationship of the present
result with Schwinger's non-local current\cite{sch} will be also
discussed. In \S\ref{model}, sum rule values of relativistic
nuclear models\cite{mc,fur,ma} will be examined, according to new insight
of the present paper. Moreover, non-relativistic sum values will be derived
from relativistic ones in the same framework. 
The final section will be devoted to a brief summary of the present work.

\section{The four-current}\label{fcurrent}

It is clear that the contradiction in relativistic sum values stems
from Eq.(\ref{field}) which is normally used for calculations
in the field theory. In our formalism, therefore, we begin with
the definition of the nucleon field,
\begin{equation}
\psi(\vct{x})= \sum_\alpha\Theta_\alpha w_\alpha(\vct{x})a_\alpha.
 \label{newfield}
\end{equation}
Here, we have used following abbreviations,
\begin{equation}
\Theta_{\vct{p}} =\theta(P_\infty-|\vct{p}|),\quad
w_\alpha(\vct{x}) = \frac{1}{\sqrt{V}}w_s(\vct{p}\sigma)
e^{i\vct{p}\cdot\vct{x}},
\quad a_\alpha = a_s(\vct{p}\sigma), 
\end{equation}
where $\alpha$ denotes $\{s=\pm,\vct{p},\sigma\}$, $V$ the volume
of the system, and $w_s$ the spinor,
\begin{align} 
w_+(\vct{p}\sigma) 
&= 
\sqrt{ \frac{E_{\vct{p}}+m}{2E_{\vct{p}}}} 
\left( 
\begin{array}{c} 
\chi_\sigma \\[4pt] 
\dfrac{\vct{\sigma}\cdot\vct{p}}{E_{\vct{p}}+m} 
\,\chi_\sigma 
\end{array} 
\right), 
\\[8pt] 
w_-(\vct{p}\sigma) 
&= 
\sqrt{ \frac{E_{\vct{p}}+m}{2E_{\vct{p}}}} 
\left( 
\begin{array}{c} 
-\dfrac{\vct{\sigma}\cdot\vct{p}}{E_{\vct{p}}+m} 
\,\chi_\sigma \\[4pt] 
\chi_\sigma 
\end{array} 
\right),
\end{align}
with $E_{\vct{p}}=\sqrt{\vct{p}^2 + m^2}$ and
the 2-component spinor, $\chi_{\sigma}$.
The notations, $a_+(\vct{p}\sigma)$ and $a_-(\vct{p}\sigma)$, stand for the
annihilation operator of a particle and the creation operator
of an antiparticle, respectively, satisfying
\begin{equation}
\{a_s(\vct{p}\sigma),\ a_{s'}^\dagger(\vct{p}'\sigma')\}
= \delta_{\vct{p}\vct{p}'}
\delta_{\sigma\sigma'}\delta_{ss'},\quad \textrm{others} = 0.
\end{equation}
In the above field, the range of $|\vct{p}|$ is restricted by $P_\infty$,
which is finite for a while. In the limit $P_\infty \rightarrow \infty$, 
Eq.(\ref{newfield}) is reduced to the usual field,
but we will take the limit later.

The anti-commutation relation of the new field becomes of 
\begin{equation}
\{\psi_m(\vct{x}), \ \psi_n^\dagger(\vct{y})\  \}
  =D_{mn}(\vct{x},\vct{y}),\label{newcom}
\end{equation}
where we have defined
\begin{align}
D_{mn}(\vct{x}, \vct{y})
&= \sum_\alpha\Theta_\alpha
\left(w_\alpha(\vct{x})w_\alpha^\dagger(\vct{y})\right)_{mn} 
= \delta_{mn}d(\vct{x}-\vct{y}),\label{D}
\\
d(\vct{x})
&= \frac{1}{V}\sum_{\vct{p}}\Theta_{\vct{p}}e^{i\vct{p}\cdot\vct{x}}
 =\int\frac{d^3p}{(2\pi)^3}\,\Theta_{\vct{p}}e^{i\vct{p}\cdot\vct{x}}.
\label{d}
\end{align}
It is seen that Eq.(\ref{newcom}) is reduced to Eq.(\ref{field})
in the limit $P_\infty \rightarrow \infty$, since
\begin{equation}
d(\vct{x}) \longrightarrow \delta(\vct{x}),\ \quad (P_\infty\rightarrow \infty).
\end{equation}

The commutation relation between currents is calculated by using the following
equation for arbitrary $4 \times 4$ matrices,
$\Gamma_1(\vct{x})$ and $\Gamma_2(\vct{x})$,
\begin{align}
[\psi^\dagger(\vct{x})\Gamma_1(\vct{x})\psi(\vct{x}),
 \ \psi^\dagger(\vct{y})\Gamma_2(\vct{y})\psi(\vct{y})] 
 &= \psi^\dagger(\vct{x})\Gamma_1(\vct{x})D(\vct{x}, \vct{y})
 \Gamma_2(\vct{y})\psi(\vct{y}) \nonumber\\[4pt]
& \phantom{=}
-\psi^\dagger(\vct{y})\Gamma_2(\vct{y})D(\vct{y}, \vct{x})
 \Gamma_1(\vct{x})\psi(\vct{x}).\label{gamma12}
\end{align}
For $\Gamma_1=1$ and $\Gamma_2=\gamma^0\vct{\gamma}$, we have
\begin{equation}
[J^0(\vct{x}), \ \vct{J}(\vct{y})]
 = \Bigl( \overline{\psi}(\vct{x})\vct{\gamma}\psi(\vct{y})
 - \overline{\psi}(\vct{y})\vct{\gamma}\psi(\vct{x}) \Bigr)
d(\vct{x}-\vct{y}),\label{cc}
\end{equation}
which does not vanish for the finite value of $P_\infty$,
differently from Eq.(\ref{current}).

Since Eq.(\ref{gamma12}) holds, even if $\Gamma$ contains differential
operators, we obtain
\begin{equation}
[ H , \ \psi^\dagger(\vct{x})\Gamma(\vct{x})\psi(\vct{x})]
 = \left(h_0(\vct{x})\psi(\vct{x})\right)^\dagger\Gamma(\vct{x})\psi(\vct{x})
- \psi^\dagger(\vct{x})\Gamma(\vct{x})h_0(\vct{x})\psi(\vct{x}),\label{H}
\end{equation}
where we have used the fact that
\[
 \int d^3y\,D(\vct{x}, \vct{y})h_0(\vct{y})\psi(\vct{y})
=h_0(\vct{x})\psi(\vct{x})
\]
for the one-body Hamiltonian,
\[
 h_0(\vct{x})=-\,i\vct{\alpha}\cdot\vct{\nabla}+\beta m\,,\qquad
H = \int d^3x\,\psi^\dagger(\vct{x})h_0(\vct{x})\psi(\vct{x}).
\]
From Eq.(\ref{H}), the equation of the current conservation is obtained,
like Eq.(\ref{currentcons}),
\begin{equation}
[H, \ J^0(\vct{x})\ ] = i\left(\vct{\nabla}\psi^\dagger(\vct{x})\right)\cdot
\vct{\alpha}\psi(\vct{x}) + i\psi^\dagger(\vct{x})\vct{\alpha}\cdot\vct{\nabla}
\psi(\vct{x}) = i\vct{\nabla}\cdot\vct{J}(\vct{x}),
\end{equation}
which gives the same expression of the double commutator as in Eq.(\ref{com}).
 
For the one-body operator,
\begin{equation}
 F_i=\int d^3x\,\psi^\dagger(\vct{x})\Gamma_i(\vct{x})\psi(\vct{x}),
 \end{equation}
we have from Eq.(\ref{gamma12})    
\begin{equation}
[F_1,\ F_2\ ]=\int d^3xd^3y\,
 \psi^\dagger(\vct{x})\Bigl(
 \Gamma_1(\vct{x})D(\vct{x}, \vct{y})
 \Gamma_2(\vct{y}) -\Gamma_2(\vct{x})D(\vct{x}, \vct{y})
 \Gamma_1(\vct{y})
\Bigr)\psi(\vct{y}),\label{gamma12'}
\end{equation}
and from Eq.(\ref{H})
\begin{equation}
[H,\ F_1\ ]
=\int d^3x\,\psi^\dagger(\vct{x})[\ h_0(\vct{x}), \Gamma_1(\vct{x})\ ]
\psi(\vct{x}).
\end{equation}
The above two equations give another expression of the double commutator
of Eq.(\ref{s}),
\begin{align}
[\ F^\dagger,[\ H, \ F\ ]\ ] 
&= \int d^3xd^3y\,
 \psi^\dagger(\vct{x})\Bigl(
 f^\ast(\vct{x})D(\vct{x}, \vct{y})
 [\ h_0(\vct{y}), f(\vct{y})\ ] \  \nonumber\\
& \hspace{2.8cm}
 - [\ h_0(\vct{x}), f(\vct{x})\ ]D(\vct{x}, \vct{y})
 f^\ast(\vct{y})
\Bigr) \psi(\vct{y}).\label{com'}
\end{align}
If we take the limit $P_\infty \rightarrow \infty$ in the above equation,
$D(\vct{x},\vct{y})$ becomes to be $\delta(\vct{x}-\vct{y})$,
so that we have the undesired result:
\begin{equation}
[\ F^\dagger,[\ H, \ F\ ]\ ] 
= \int d^3x\,
 \psi^\dagger(\vct{x})[\ f^\ast(\vct{x}),\ [\ h_0(\vct{x}),\  f(\vct{x})\ ] \ ]
 \psi(\vct{x}) = 0,\label{com''x}
\end{equation}
as mentioned in \S\ref{intro}. It will be shown that the limit
$P_\infty \rightarrow \infty$ should be taken, after calculating
the ground-state expectation value of Eq.(\ref{com'}) or Eq.(\ref{cc})
used in Eq.(\ref{com}).

\section{Expectation value of the current}\label{exp}

Let us calculate the ground-state expectation value of Eq.(\ref{cc}).
Assuming isospin symmetric nuclear matter with Fermi momentum $\kf$,
we have 
\begin{equation}
\mtrix{0}{\overline{\psi}(\vct{x})\vct{\gamma}\psi(\vct{y})}{0}
 =\frac{4}{V}\sum_p\left(\theta_{\vct{p}} - \Theta_{\vct{p}}\right)
 \frac{\vct{p}}{E_{\vct{p}}}
 e^{-i\vct{p}\cdot(\vct{x}-\vct{y})} = \vct{j}(\vct{x} - \vct{y}),
\end{equation}
where the notation,
$\theta_{\vct{p}} = \theta(\kf - |\vct{p}|)$, is used.
The term with $\Theta_{\vct{p}}$ comes from contributions of the Dirac sea.
Using the above equation, the expectation value of Eq.(\ref{cc}) is
expressed as
\begin{equation}
\mtrix{0}{[iJ^0(\vct{x}),\ \vct{J}(\vct{y})]}{0}
 = 2i\vct{j}(\vct{x} - \vct{y})d(\vct{x} - \vct{y})
 = \vct{K}(\vct{x} - \vct{y})\label{K''}
\end{equation}
with 
\begin{align} 
\vct{K}(\vct{x}) 
&= \frac{8i}{V^2}\sum_{\vct{p},\vct{q}} 
\left(\theta_{\vct{p}} - \Theta_{\vct{p}}\right)\Theta_{\vct{p}+\vct{q}} 
\frac{\vct{p}}{E_{\vct{p}}}e^{i\vct{q}\cdot\vct{x}} \nonumber \\ 
&= 
\frac{8i}{V^2}\sum_{\vct{p},\vct{q}} 
\left(\theta_{\vct{p}} - \Theta_{\vct{p}}\right)\Theta_{\vct{p}+\vct{q}} 
\frac{\vct{q}\,\vct{p}\cdot\vct{q}/q^2}{E_{\vct{p}}}e^{i\vct{q}\cdot\vct{x}} 
\nonumber \\ 
&= 
\frac{2i}{V^2}\sum_{\vct{q}}e^{i\vct{q}\cdot\vct{x}}\frac{\vct{q}} 
{q^2}\Bigl( \snn(\vct{q}) + \snbar(\vct{q}) \Bigr),\label{K'} 
\end{align} 
where $\snn(\vct{q})$ and $\snbar(\vct{q})$ stand for
\begin{equation}
\snn(\vct{q}) = 4\sum_{\vct{p}}
\theta_{\vct{p}}\Theta_{\vct{p}+\vct{q}}
\frac{\vct{p}\cdot\vct{q}}{E_{\vct{p}}},
\qquad
\snbar(\vct{q}) = -4\sum_{\vct{p}}
\Theta_{\vct{p}}\Theta_{\vct{p}+\vct{q}}
\frac{\vct{p}\cdot\vct{q}}{E_{\vct{p}}}.
\label{SS}
\end{equation}
The calculation of $\snn(\vct{q})$ is performed by replacing
the sum with the integral.
Defining $t=\vct{p}\cdot\vct{q}/pq$, $\snn(\vct{q})$ is expressed as
\begin{equation}
\snn(\vct{q})
= \frac{Vq}{\pi^2}\int_0^{\kf}dp\,\frac{p^3}{E_p}\int_{-1}^1dt\,
\Theta_{\vct{p}+\vct{q}}t=\frac{Vq}{\pi^2}
\int^{\kf}_0 dp\,\frac{p^3}{E_p}\int_{-1}^1dt\,\theta(\xi-t)t,
\end{equation} 
with
\[
\xi=\frac{P_\infty^2-p^2-q^2}{2pq},
\]
where $\xi$ should be $\xi> -1$. Finally we have
\begin{equation}
\snn(\vct{q})= \left\{
\begin{array}{ll}
0, & |P_\infty - q| > \kf ;\\[6pt]
\dfrac{V}{8\pi^2q}G(|P_\infty - q|, \ \kf), & |P_\infty - q| < \kf,
\end{array}
\right.
\label{sn}
\end{equation}
where $G$ is defined as
\begin{equation}
G(a, b) = \int_a^bdp\,\frac{p}{E_p}
\left((p+q)^2 - P_\infty^2\right)\left((p-q)^2 - P_\infty^2\right).
\end{equation}
The above result will be understood as follows.
The value of $\snn(\vct{q})$ vanishes
in the region $q < P_\infty - \kf$ where $\Theta_{\vct{p}+\vct{q}} =1$,
since the sum of $\vct{p}$ cancels the terms
with $\vct{p}$ and $-\vct{p}$ in Eq.(\ref{SS}),
while in the region $|P_\infty - q| < \kf$,
there is no such perfect cancellation in the sum of $\vct{p}$.
In the region $q > P_\infty + \kf$,
$\Theta_{\vct{p}+\vct{q}} =0$ yields $\snn(\vct{q})=0$.

The expression of $\snbar(\vct{q})$ from the Dirac sea is obtained
by replacing $\kf$ in Eq.(\ref{sn}) with $P_\infty$,
\begin{equation}
\snbar(\vct{q})
= \left\{
\begin{array}{ll}
0,  &  q > 2P_\infty; \\[6pt]
-\dfrac{V}{8\pi^2q}G(|P_\infty - q|, \  P_\infty), & q < 2P_\infty.
\end{array}\right.
\label{snbar}
\end{equation}
In the region $q > 2P_\infty$, the value of $\snbar(\vct{q})$
disappears, because of $\Theta_{\vct{p}}\Theta_{\vct{p}+\vct{q}}=0$
in Eq.(\ref{SS}).

If there is not the factor with $P_\infty$ in Eq.(\ref{SS}), both $\snn$
and $\snbar$ vanish. The existence of $P_\infty$ yields a constraint
on the states which contribute to Eq.(\ref{SS}). Thus, the operation
of $P_\infty \rightarrow \infty$ and the calculation of Eq.(\ref{SS})
do not commute with each other.

The region which we may be physically interested in is
in a range $q\ll P_\infty$ for $m\ll P_\infty$.
In this region, the value of $\snn(\vct{q})$ vanishes,
but the one of $\snbar(\vct{q})$ does not.
When we expand the function $G$ in Eq.(\ref{snbar})
in terms of $q/P_\infty$ for $m\ll P_\infty$, we have
\begin{align}
 G(|P_\infty - q|, \  P_\infty)
&=  P_{\infty}^5\int_0^{q/P_\infty}dx
  \frac{1-x}{\sqrt{(1-x)^2+(m/P_\infty)^2}} \notag \\
&\phantom{=P^5}
 \times\left((x^2-2x)^2-2(x^2-2x+2)\frac{q^2}{P_\infty^2}
	    + \frac{q^4}{P_{\infty}^4}\right)\notag \\
&\approx
 -\frac{8}{3}P_{\infty}^5\left(1-\frac{m^2}{2P_{\infty}^2}\right)
\frac{q^3}{P_{\infty}^3}
\left(1-\frac{3q}{8P_\infty}-\frac{q^2}{5P_{\infty}^2}\right),
\label{g}
\end{align}
where $x$ is defined by $p=P_\infty(1-x)$. The above equation shows
that when $P_\infty \rightarrow \infty$,
the value of $\snbar(\vct{q})$ is divergent,
\begin{equation}
\snbar(\vct{q}) = \frac{V}{3\pi^2}q^2P^2_\infty.
\label{div}
\end{equation}

More intuitive derivation of Eq.(\ref{div}) may be performed
by expanding the step function
$\Theta_{\vct{p}+\vct{q}}$ near $|\vct{p}|\approx P_\infty$ in Eq.(\ref{SS}), 
\begin{equation}
\Theta_{\vct{p}+\vct{q}}
= \theta(P_\infty-|\vct{p}|-\Delta p)
= \Theta_{\vct{p}}
- \Delta p\,\delta(P_\infty-|\vct{p}|)+\cdots,\quad
\Delta p = |\vct{p}+\vct{q}|-|\vct{p}|
\end{equation}
which yields
\begin{equation}
\Theta_{\vct{p}}\Theta_{\vct{p}+\vct{q}}
=\Theta_{\vct{p}}-\frac{\Delta p}{2}\delta(P_\infty-|\vct{p}|)
 +\cdots .
\end{equation}
This result together with Eq.(\ref{SS}) provides us with
\begin{equation}
\snbar(\vct{q})
\approx 2\sum_{\vct{p}}\Delta p\, \delta(P_\infty-|\vct{p}|) 
\frac{\vct{p}\cdot\vct{q}}{E_{\vct{p}}}
\approx  \frac{V}{3\pi^2}q^2P^2_\infty,
\label{div2}
\end{equation}
for $m\ll P_\infty$, as in Eq.(\ref{div}).

In the limit  $P_\infty \rightarrow \infty$ for $q\ll P_\infty$
where $\snn(\vct{q})$ vanishes and $\snbar(\vct{q})$
is given by Eq.(\ref{div2}), Eq.(\ref{K'}) is described as
\begin{equation}
\vct{K}(\vct{x})
= \frac{2P^2_\infty}{3\pi^2V}\vct{\nabla}\sum_{\vct{q}}
e^{i\vct{q}\cdot\vct{x}}
= \frac{2P^2_\infty}{3\pi^2}\vct{\nabla}\delta(\vct{x}).
\end{equation}
Consequently, we obtain the expression for the commutator of the currents,
\begin{equation}
\mtrix{0}{[iJ^0(\vct{x}),\ \vct{J}(\vct{y})]}{0}
 =
\frac{2P^2_\infty}{3\pi^2}\vct{\nabla}_{\vct{x}}\delta(\vct{x}-\vct{y}),
\label{newcom'}
\end{equation}
This is nothing but the gradient of $\delta$-function required by Schwinger. 

Instead of Eq.(\ref{fourcurrent}), Schwinger\cite{sch} assumed
the point-split current for the space part,  
\begin{equation}
J^{\mu}_\epsilon(\vct{x})
= \overline{\psi}(\vct{x}-\vct{\epsilon}/2)
\gamma^\mu\psi(\vct{x}+\vct{\epsilon}/2)
\quad (\ \vct{\epsilon} \rightarrow 0 \ ),\label{local}
\end{equation}
which gives the commutation relation with the time-component,
\begin{align}
[ \ J^0(\vct{x}),  \  \vct{J}_\epsilon(\vct{y})\ ]
&=\Bigl(
\delta(\vct{x}-\vct{y}+\vct{\epsilon}/2)
-\delta(\vct{x}-\vct{y}-\vct{\epsilon}/2)
\Bigr)
\vct{J}_\epsilon(\vct{y}) \label{scu} \\
&=
 \vct{J}_\epsilon(\vct{y})\vct{\epsilon}
 \cdot\vct{\nabla}_{\vct{x}}\delta(\vct{x}-\vct{y}). \nonumber
\end{align}
The calculation of the ground-state expectation value 
provides us with\cite{sch},
\begin{equation}
\mtrix{0}{i[J^0(\vct{x}),\ \vct{J}(\vct{y})]}{0}
=
\frac{4}{3\pi^2\epsilon^2}\vct{\nabla}_{\vct{x}}\delta(\vct{x}-\vct{y}).
\label{scom}
\end{equation}
This has divergent limit of $1/\epsilon^2$, and
is the same as the present result Eq.(\ref{newcom'}),
when setting $P^2_\infty = 2/\epsilon^2$.

The relationship between the present current and Schwinger's one
may be also seen qualitatively  as follows.
On the one hand, Eq.(\ref{scu}) is written as
\begin{equation}
[ \ J^0(\vct{x}),  \  \vct{J}_\epsilon(\vct{y})]
 =\overline{\psi}(\vct{x})\vct{\gamma}\psi(\vct{x}+\vct{\epsilon})
\delta(\vct{x}-\vct{y}+\vct{\epsilon}/2)
-
\overline{\psi}(\vct{x}-\vct{\epsilon})\vct{\gamma}
\psi(\vct{x})\delta(\vct{x}-\vct{y}-\vct{\epsilon}/2).
\label{scu''}
\end{equation}
On the other hand, the present function $d(\vct{x})$ in Eq.(\ref{d})
is calculated as
\begin{equation}
d(\vct{x})=-\frac{1}{2\pi^2x}\frac{d\ }{dx}\frac{\sin{P_\infty x}}{x},
\end{equation}
which has the spreading width about $|x|\lesssim 2\pi/P_\infty$ around $x=0$.
Therefore, if we replace $\psi(\vct{y})d(\vct{x}-\vct{y})$ by
$\psi(\vct{x}\pm\vct{\epsilon})\delta(\vct{x}-\vct{y}\pm\vct{\epsilon}/2 )$
with $\epsilon \sim 1/ P_\infty$ in Eq.(\ref{cc}),
we have the same form as Eq.(\ref{scu''}) of Schwinger's model.

Thus, the commutator with the non-local current Eq.(\ref{local})
assumed by Schwinger\cite{sch} is well understood as a fact that
the limit $P_\infty \rightarrow \infty$ should be taken after
calculating the expectation value
of the commutator in the local field theory.

\section{Sum values in relativistic nuclear models}\label{model}

The sum value $S$ of the excitation strengths for the operator
$f(\vct{x})$ is obtained by Eq.(\ref{com}) and Eq.(\ref{newcom'}) as,
\begin{equation}
S=\frac{1}{2}\mtrix{0}{[\ F^\dagger,\ [\ H, \ F\ ]\ ]}{0}
=\frac{P^2_\infty}{3\pi^2}\int d^3x\,
|\vct{\nabla}f(\vct{x})|^2.\label{sumvalue1}
\end{equation}
In writing the excitation operator in a momentum space,
\begin{equation}
f(\vct{x})=\sum_{\vct{q}}f(\vct{q})e^{i\vct{q}\cdot\vct{x}},
\quad f(\vct{q})= \frac{1}{V}
\int d^3x\, f(\vct{x})e^{-i\vct{q}\cdot\vct{x}},
\end{equation}
Eqs.(\ref{K''}) and (\ref{K'}) give the sum value of the form:
\begin{equation}
S
=
\frac{1}{2} \int d^3xd^3y\, f^\ast(\vct{x})f(\vct{y})
\vct{\nabla}_y\cdot \vct{K}(\vct{x}-\vct{y})
=
\sum_{\vct{q}}|f(\vct{q})|^2
\Bigl( \snn(\vct{q})+\snbar(\vct{q}) \Bigr) .
\label{sumvalue2} 
\end{equation}  
Eq.(\ref{sumvalue1}) is for the case of
$m, q\ll P_\infty\ (P_\infty \rightarrow \infty)$,
where $\snn(\vct{q})=0$, in the above equation.

When the function $f(\vct{x})$ is given by Eq.(\ref{fx}),
Eq.(\ref{sumvalue2}) becomes of
\begin{equation}
S=\snn(\vct{q})+\snbar(\vct{q}).\label{sumvalue3} 
\end{equation}
It will be shown later how the above equation is reduced
to the nonrelativistic f-sum rule value Eq.(\ref{fsum}).

According to the result Eq.(\ref{sumvalue2}), let us discuss the sum
values of relativistic nuclear models which have been extensively used
for nuclear study, and shown to work well
phenomenologically\hspace{0pt}\cite{sw,ring}.

The full energy-weighted transition strengths of $A$ nucleon system
in the mean field are given by 
\begin{equation}
S = \sph +\spnbar \label{ts}
\end{equation}
with
\[
\sph = \sum_{ph}(E_p-E_h)|\mtrix{p}{f}{h}|^2,
\quad 
\spnbar = \sum_{b<0,p}(E_p-E_b)|\mtrix{p}{f}{b}|^2,
\]
where $\{\textrm{ph}\}$ and $\{\textrm{p}\bar{\textrm{N}}\}$
represent particle-hole
and particle-antinucleon excitations, respectively, and $\ket{b}\  (b<0)$
stands for the negative-energy states.
The second term $\spnbar$ is described as
\begin{equation}
\spnbar = \svac - S_{\textrm{Pauli}},
\end{equation}
where $\svac$ denotes the transitions of
antinucleons in the vacuum, that is, Dirac sea, to the
positive-energy states, and $S_{\textrm{Pauli}}$ the Pauli
blocking terms due to the existence of $A$ nucleons,
\begin{equation}
\svac = \sum_{b<0,a>0}(E_a-E_b)|\mtrix{a}{f}{b}|^2, \quad
S_{\textrm{Pauli}}=\sum_{b<0,h}(E_h-E_b)|\mtrix{h}{f}{b}|^2.
\end{equation}
Then, the total sum $S$ in Eq.(\ref{ts}) is written as, 
\begin{equation}
S = \snosea + \svac, \label{ts'}
\end{equation}
where $\snosea$ is defined by
\begin{equation}
\snosea = \sph - S_{\textrm{Pauli}} = \sph + \snbarh.
\end{equation}
Thus, $\snosea$ is nothing but the sum value in the no-sea
approximation which is extensively used in relativistic nuclear
models\cite{fur,ma}. In this approximation,
the negative-energy states are assumed to be empty,
$\svac=0$. As a result, there is the second
term $\snbarh$ which
represents transitions of the particles in the Fermi sea to
the negative-energy states with negative excitation
energies, yielding unphysical negative energy-weighted strengths
in \=N excitation energy region as a cost of neglecting
$\svac$\cite{kura5}.

Since we have the identities,
\begin{equation}
\sum_{h,h'}(E_{h'}-E_h)|\mtrix{h'}{f}{h}|^2=0,\quad
\sum_{b<0,b'<0}(E_{b'}-E_b)|\mtrix{b'}{f}{b}|^2=0,
\end{equation}
we can write $\snosea$ and $\svac$ as
\begin{equation}
\begin{split}
\snosea 
&= \sum_{h,\alpha}(E_{\alpha}-E_h)|\mtrix{\alpha}{f}{h}|^2
 =\sum_{h,\alpha}\mtrix{h}{f^\ast}{\alpha}\mtrix{\alpha}{[\ h_0,\ f\ ]}{h},
\\ 
\svac
&= \sum_{b<0,\alpha}(E_{\alpha}-E_b)|\mtrix{\alpha}{f}{b}|^2
 =\sum_{b<0,\alpha}\mtrix{b}{f^\ast}{\alpha}\mtrix{\alpha}{[\ h_0,\ f\ ]}{b},
\end{split}
\label{swsum}
\end{equation}
where $\alpha$ denotes both positive $\ket{a}\  (a>0) $ and
negative $\ket{b}\  (b<0)$ energy states .
When we express the time-reversal state of $\ket{a}$ by
$\ket{\bar a}$, they are also written as
\begin{align*}
\snosea 
&=
\sum_{h,\alpha}
\mtrix{\bar h}{f}{\bar\alpha}
\mtrix{\bar\alpha}{[\ h_0,\ f^\ast\ ]}{\bar h}
=
-\sum_{h,\alpha}
\mtrix{h}{[\ h_0,\ f\ ]}{\alpha}\mtrix{\alpha}{f^\ast}{h}, \\
\svac
&=
\sum_{b<0,\alpha}\mtrix{\bar b}{f}{\bar\alpha}
\mtrix{\bar\alpha}{[\ h_0,\ f^\ast\ ]}{\bar b}
= -\sum_{b<0,\alpha}
\mtrix{b}{[\ h_0,\ f\ ]}{\alpha}\mtrix{\alpha}{f^\ast}{b},
\end{align*}
which give the expressions;
\begin{align*}
\snosea
&=
\frac{1}{2}\sum_{h,\alpha}
\Bigl( 
\mtrix{h}{f^\ast}{\alpha}\mtrix{\alpha}{[\ h_0,\ f\ ]}{h}
-\mtrix{h}{[\ h_0,\ f\ ]}{\alpha}
\mtrix{\alpha}{f^\ast}{h}
\Bigr), \\
\svac
&=
 \frac{1}{2}\sum_{b<0,\alpha}
\Bigl(
\mtrix{b}{f^\ast}{\alpha}\mtrix{\alpha}{[h_0,f]}{b}
-\mtrix{b}{[\ h_0,\ f\ ]}{\alpha}
\mtrix{\alpha}{f^\ast}{b}
\Bigr).
\end{align*}
Now, if we were able to use the closure property in the intermediate states
$\sum_\alpha\ket{\alpha}\bra{\alpha}=1$, we would have
\begin{equation}
\snosea
=\sum_{h}\mtrix{h}{[\ f^\ast,\ [\ h_0,f\ ]\ ]}{h}=0,\quad 
\svac
=\sum_{b<0}\mtrix{b}{[\ f^\ast,\ [\ h_0,\ f\ ]\ ]}{b}=0,
\end{equation}
which led to misunderstanding the relativistic sum values\cite{mc}.
We can not use the closure property, since there are $\ket{\alpha}$
states which should be excluded by the step function with $P_\infty$.
In fact, we have to calculate Eq.(\ref{swsum}) as follows,
\begin{align*}
\snosea
&=
\sum_{h,\alpha}\theta_h\Theta_{\alpha}\int d^3xd^3y\, f^\ast(\vct{x})
w_h^\dagger(\vct{x})w_{\alpha}(\vct{x})w_{\alpha}^\dagger(\vct{y})
\bigl(-i\vct{\alpha}\cdot\vct{\nabla}f(\vct{y})\bigr)w_h(\vct{y}) \\
&=
-\frac{2i}{V}\sum_{\vct{p}}\theta_{\vct{p}}
\int d^3xd^3y\, e^{-i\vct{p}\cdot(\vct{x}-\vct{y})}
d(\vct{x}-\vct{y})f^\ast(\vct{x})\vct{\nabla}f(\vct{y})\cdot
\textrm{Tr}(\vct{\gamma}\Lambda_{\vct{p}}^+), \\
\svac
&=
\sum_{b<0,\alpha}\Theta_b\Theta_{\alpha}
\int d^3xd^3y\, f^\ast(\vct{x})w_{b}^\dagger(\vct{x})
w_{\alpha}(\vct{x})w_{\alpha}^\dagger(\vct{y})
\bigl(-i\vct{\alpha}\cdot\vct{\nabla}f(\vct{y})\bigr)w_{b}(\vct{y}) \\
&=
 - \frac{2i}{V}\sum_{\vct{p}}\Theta_{\vct{p}}\int d^3xd^3y\,
e^{-i\vct{p}\cdot(\vct{x}-\vct{y})}
d(\vct{x}-\vct{y})f^\ast(\vct{x})\vct{\nabla}f(\vct{y})\cdot
\textrm{Tr}(\vct{\gamma}\Lambda_{\vct{p}}^-),
\end{align*}
where the projection operator $\Lambda^{\pm}_{\vct{p}}$ is defined as 
\begin{equation}
\Lambda^{\pm}_{\vct{p}}
=
\sum_{\sigma}w_{\pm}(\vct{p}\sigma)\overline{w}_{\pm}(\vct{p}\sigma)
=
\frac{E_{\vct{p}}\gamma^0 \mp \vct{p}\cdot\vct{\gamma} \pm m }{2E_{\vct{p}}}.
\end{equation}
By using the expressions of $d(\vct{x})$ and $f(\vct{x})$ in momentum space,
and the fact that
Tr($\vct{\gamma}\Lambda_{\vct{p}}^\pm)=\pm2\vct{p}/E_{\vct{p}}$,
finally we obtain
\begin{equation}
\snosea
=
4\sum_{\vct{p},\vct{q}}\theta_{\vct{p}}\Theta_{\vct{p}+\vct{q}}|f(\vct{q})|^2
\frac{\vct{p}\cdot\vct{q}}{E_{\vct{p}}}
=
\sum_{\vct{q}}|f(\vct{q})|^2\snn(\vct{q}),\quad
\svac
=\sum_{\vct{q}}|f(\vct{q})|^2 \snbar(\vct{q}).
\end{equation}

When $q\ll P_\infty$, we can replace
$\theta_{\vct{p}}\Theta_{\vct{p}+\vct{q}}$ by $\theta_{\vct{p}}$
in the above $\snn(\vct{q})$. Therefore, as far as discussions on
$\snosea$ are concerned, we can set
$P_\infty \rightarrow \infty$ at the beginning of calculations,
which gives $[\ f^\ast,\ [\ h_0,\ f\ ]\ ]=0$, and $\snosea=0$.

The sum value of $\sph$ is given by,
\begin{align*}
\sph
&=
\sum_{a>0,h}\mtrix{h}{f^\ast}{a}\mtrix{a}{[\ h_0,\ f\ ]}{h} \\
&=
\sum_{a>0,h}\theta_h\Theta_a\int d^3xd^3y\, f^\ast(\vct{x})
 w_h^\dagger(\vct{x})w_a(\vct{x})w_a^\dagger(\vct{y})
 \left(-i\vct{\alpha}\cdot\vct{\nabla}f(\vct{y})\right)w_h(\vct{y}) \\
&=
-\frac{2i}{V}\sum_{\vct{p}}\theta_{\vct{p}}
\int d^3xd^3y\,e^{-i\vct{p}\cdot(\vct{x}-\vct{y})}
f^\ast(\vct{x})\vct{\nabla}f(\vct{y})\cdot
\textrm{Tr}
\Bigl(
D^+(\vct{x},\vct{y})\vct{\alpha}\Lambda_{\vct{p}}^+\gamma_0
\Bigr),
\end{align*}
where $D^+(\vct{x},\vct{y})$ is defined, as in Eq.(\ref{D}), with
\begin{equation}
D^+(\vct{x}, \vct{y})
=
\sum_{\vct{p}\sigma}\Theta_{\vct{p}}
w_{+\vct{p}\sigma}(\vct{x})w_{+\vct{p}\sigma}^\dagger(\vct{y})
=
\frac{1}{V}\sum_{\vct{p}}\Theta_{\vct{p}}e^{i\vct{p}\cdot(\vct{x}-\vct{y})}
\Lambda_{\vct{p}}^+\gamma_0.
\end{equation}
It is calculated to be
\begin{equation}
\sph=\sum_{\vct{q}}|f(\vct{q})|^2 \sph(\vct{q}), \quad
\sph(\vct{q})
 =2\sum_{\vct{p}}\theta_{\vct{p}}\Theta_{\vct{p}+\vct{q}}
\left(
\frac{(\vct{p}+\vct{q})\cdot\vct{q}}{E_{\vct{p}+\vct{q}}}
+\frac{\vct{p}\cdot\vct{q}}{E_{\vct{p}}}
\right).\label{phsq}
\end{equation}

The transitions of particles in Fermi sea to negative energy
states are calculated in the same way,
\begin{equation}
\snbarh
=
\sum_{\vct{q}}|f(\vct{q})|^2 \snbarh(\vct{q}),\quad
\snbarh(\vct{q})
=
-2\sum_{\vct{p}}\theta_{\vct{p}}\Theta_{\vct{p}+\vct{q}}
\left(
\frac{(\vct{p}+\vct{q})\cdot\vct{q}}{E_{\vct{p}+\vct{q}}}
 -\frac{\vct{p}\cdot\vct{q}}{E_{\vct{p}}}
\right).\label{nbarhsq}
\end{equation}
The sum of $\sph$ in Eq.(\ref{phsq}) and
$\snbarh(\vct{q})$ in Eq.(\ref{nbarhsq})
gives $\snn(\vct{q})$ as
\begin{equation}
\snn(\vct{q})
=4\sum_{\vct{p}}\theta_{\vct{p}}
\Theta_{\vct{p}+\vct{q}}\frac{\vct{p}\cdot\vct{q}}{E_{\vct{p}}}
=0
\end{equation}
for $\theta_{\vct{p}}\Theta_{\vct{p}+\vct{q}}
=\theta_{\vct{p}} \ \  (P_\infty-q > \kf)$, as seen in Eq.(\ref{sn}).
 
In nonrelativistic approximation, we may replace $E_{\vct{p}+\vct{q}}$
and $E_{\vct{p}}$ by $m$
in Eq.(\ref{phsq}), so that we obtain
 \begin{equation}
\sph^{\textrm{NR}}
= \sum_{\vct{q}}|f(\vct{q})|^2S_0(q)
= \frac{\rho}{2m}\int d^3x\,|\vct{\nabla}f(\vct{x})|^2,\quad
\rho=\frac{A}{V}=\frac{2\kf^2}{3\pi^2},\label{nlimit}
\end{equation}
where $S_0(q)$ is defined in the limit $q \rightarrow 0$ as,
\begin{equation}
\sph(\vct{q}) \rightarrow
S_0(q)= \frac{A}{2E_{\textrm{F}}}q^2,
\quad E_{\textrm{F}}= \sqrt{\kf^2+m^2}.
\end{equation}
The above $\sph^{\textrm{NR}}$ is the form of Eq.(\ref{sff})
for nuclear matter, and gives the f-sum rule
Eq.(\ref{fsum}) in the nonrelativistic framework.

As $q \rightarrow 0$, $\sph(\vct{q})$ in Eq.(\ref{phsq}) is
proportional to $q^2$ like the f-sum rule,
while as $q \rightarrow \infty$, it is to $q$,
\begin{equation}
\sph(\vct{q}) \rightarrow
\frac{\kf^3}{3\pi^2}Vq\left(1-\frac{m^2+2\kf^2/5}{2q^2}\right).
\end{equation}

It should be noted that in relativistic models, $\sph$
which is reduced to the nonrelativistic sum rule value is exactly
canceled  by the subtraction of the Pauli blocking terms,
when $P_\infty-q > \kf$.
Then, the relativistic sum value $S=\svac$ stems from
the transitions of antiparticles in Dirac sea to positive energy
states, which is infinite  and is independent of the Fermi momentum
or $A$ of the nuclear system.
  
Before closing the present section, it may be useful to describe
nonrelativistic sum rules in terms of the field theory developed in
this paper. The nonrelativistic field is written as
\begin{equation}
\psi(\vct{x})
= \frac{1}{\sqrt{V}}\sum_{\vct{p}\sigma}\Theta_{\vct{p}}
\chi_{\sigma}e^{i\vct{p}\cdot\vct{x}} a(\vct{p}\sigma),
\end{equation}
which satisfies the commutation relation
$\{\psi_m(\vct{x}),\psi_n^\dagger(\vct{y})\} =\delta_{mn}d(\vct{x}-\vct{y})$.
The nuclear four-current is given by
\begin{equation}
J_0(\vct{x})=\psi^\dagger(\vct{x})\psi(\vct{x}), \quad
\vct{J}(\vct{x})
= -\frac{i}{2m}\left(\psi^\dagger(\vct{x})\vct{\nabla}\psi(\vct{x})
  -(\vct{\nabla}\psi^\dagger(\vct{x}))\psi(\vct{x})\right).
\end{equation}
Their commutator is calculated as
\begin{equation}
[iJ_0(\vct{x}), \ \vct{J}(\vct{x})]
 =\frac{1}{m}\Bigl(
d(\vct{x}-\vct{y})\vct{\nabla}_{\vct{y}}\rho(\vct{x},\vct{y})
-\rho(\vct{x},\vct{y})\vct{\nabla}_{\vct{y}}d(\vct{x}-\vct{y})
\Bigr),\label{nonc}
\end{equation}
where we have defined
\begin{equation}
\rho(\vct{x},\vct{y})
=\frac{1}{2}\left(
\psi^\dagger(\vct{x})\psi(\vct{y})+\psi^\dagger(\vct{y})\psi(\vct{x})
\right).
\end{equation}
Since we have
\begin{equation}
n(\vct{x}-\vct{y})
= \mtrix{0}{\rho(\vct{x},\vct{y})}{0}
=\frac{4}{V}
\sum_{\vct{p}}\theta_{\vct{p}}e^{-i\vct{p}\cdot(\vct{x}-\vct{y})},
\end{equation}
the expectation value of Eq.(\ref{nonc}) is described as
\begin{equation}
\mtrix{0}{[iJ_0(\vct{x}), \ \vct{J}(\vct{x})]}{0}
=\frac{1}{m}\vct{K}(\vct{x}-\vct{y})
\end{equation}
with
\begin{equation}
\vct{K}(\vct{x})
=n(\vct{x})\vct{\nabla}d(\vct{x})-d(\vct{x})\vct{\nabla}n(\vct{x})
=\frac{4i}{V^2}\sum_{\vct{p},\vct{q}}\theta_{\vct{p}}\Theta_{\vct{p}+\vct{q}}
(2\vct{p}+\vct{q})e^{i\vct{q}\cdot\vct{x}}.
\end{equation}
In the case of $P_\infty - q > \kf$, we can put
$\theta_{\vct{p}}\Theta_{\vct{p}+\vct{q}}=\theta_{\vct{p}}$
and $P_\infty=\infty$ in the above equation. Therefore, we obtain
\begin{equation}
\vct{K}(\vct{x})=\rho\vct{\nabla}\delta(\vct{x}),
\end{equation}
which provides us with Eq.(\ref{eclcurrent}) for nuclear matter;
\begin{equation}
\mtrix{0}{i[J^0(\vct{x}),\ \vct{J}(\vct{y})]}{0}= \frac{\rho}{m}
\vct{\nabla}_{\vct{x}} \delta(\vct{x}-\vct{y}).\label{eclcurrent'}
\end{equation}

\section{Conclusions}\label{con}

It is known in the relativistic field theory that
a naive equal-time commutator between the time- and space-components
of the local four-current vanishes, and that this fact leads to
the paradoxical contradiction on the linearly energy-weighted sum
of the excitation strengths\cite{goto,sch,itz,jac}.
The sum should be positive definite, but its expression
in terms of the current disappears. In order to avoid this problem, 
Schwinger\cite{sch} introduced the non-locality in the space part of the
current from Lorentz covariance considerations\cite{jac}.
The commutator of the non-local current with the time-component yields
so-called Schwinger term, which has been shown to play an important
role in relativistic field algebras\cite{itz,jac,g,w,na},
although its form is not fixed model-independently\cite{sch,jac,g} .

On the other hand,
Walecka et al.\cite{sw} proposed a relativistic nuclear model,
where nucleons are assumed to be Dirac particles. It has been shown that
nuclear response functions are well reproduced phenomenologically
by the relativistic model\cite{ma}, but that its energy-weighted sum value
is not well defined, since the double commutator of the excitation operator
with Dirac Hamiltonian vanishes\cite{mc}. The role of the Schwinger term
in this model has not been discussed so far.

In the present paper, it has been shown that
the ground state expectation value of the
commutator with Schwinger's nonlocal current is derived in an analytic way
using the local current which is defined in the finite momentum space.
By making the momentum space infinite after calculating the
expectation value, the contradiction on the energy-weighted sum
and a naive current commutator is solved. 
It has been also discussed why calculations of the expectation values
cannot be exchanged with taking the infinite momentum space.

According to the same framework as the one for Schwinger term,
the sum values of the relativistic nuclear models\cite{sw,ring}
have been examined.
It has been shown that the vanishing double commutator of
the excitation operator with Dirac Hamiltonian can be used only
in the no-sea approximation\cite{fur,ma}
where Dirac sea is assumed to be empty, but should not be used for
discussions of the total energy-weighted sum of relativistic nuclear models.

The RPA sum rules in the relativistic nuclear models
will be discussed elsewhere.

\end{document}